\documentclass[aps,  pra, twocolumn, superscriptaddress]{revtex4-1}
\usepackage{graphicx, subfigure}
\usepackage{amsmath,amssymb,bbold}
\usepackage{color}
\usepackage{natbib,hyperref}
\usepackage{placeins}
\usepackage[pagewise]{lineno}

\begin{document}

\title{Optimizing EPR pulses for broadband excitation and refocusing}
\author{Eric R. Lowe}
\email{elowe2@umbc.edu}
\affiliation{Department of Physics, University of Maryland Baltimore County, Baltimore, MD 21250, USA}
\author{Stefan Stoll}
\email{stst@uw.edu}
\affiliation{Department of Chemistry, University of Washington, Seattle, WA 98195, USA}
\author{J.~P.~Kestner}
\email{jkestner@umbc.edu}
\affiliation{Department of Physics, University of Maryland Baltimore County, Baltimore, MD 21250, USA}
\email[]{stst@uw.edu, jkestner@umbc.edu}
\begin{abstract}
In this paper, we numerically optimize broadband pulse shapes that maximize Hahn echo amplitudes. Pulses are parameterized as neural networks (NN), nonlinear amplitude limited Fourier series (FS), and discrete time series (DT). These are compared to an optimized choice of the conventional hyperbolic secant (HS) pulse shape. A power constraint is included, as are realistic shape distortions due to power amplifier nonlinearity and the transfer function of the microwave resonator. We find that the NN, FS, and DT parameterizations perform equivalently, offer improvements over the best HS pulses, and contain a large number of equivalent optimal maxima, implying the flexibility to include further constraints or optimization goals in future designs.
\end{abstract}

\maketitle


\section{Introduction}\label{sec:intro}

The use of shaped pulses in electron paramagnetic resonance (EPR) spectroscopy is a topic of recent interest \cite{prisner_pulsed_2001,doll_adiabatic_2013,spindler_perspectives_2017,doll_wideband_2017,tait_endor_2017,prisner_shaping_2019}. They address the basic challenge that the excitation bandwidth of monochromatic square pulses is much smaller than the spectral line width of samples. This situation can arise in nuclear magnetic resonance (NMR), but it is more broadly relevant in EPR. Typical EPR spectral widths are about 250 MHz for a nitroxide at Q-band ($\approx$ 1.2 T) or $\approx$ 2 GHz for a Cu(II) complex at X-band ($\approx$ 0.35 T). Using shaped pulses can increase sensitivity and excitation bandwidth.

Broadband pulses have been initially designed for NMR, including the Kunz--B\"{o}hlen--Bodenhausen (KBB) approach to generate a Hahn echo  \cite{kunz_use_1986,kunz_frequency-modulated_1987,bohlen_refocusing_1989}. This sequence consists of a frequency-swept (chirped) $\pi/2$ pulse of length $t_\mathrm{p}$, followed by a chirped $\pi$ pulse of length $t_\mathrm{p}/2$. The intention of this sequence is to refocus all the spins within a broad excitation window. The Fourier transform of this echo gives the spectral distribution of the excited spin ensemble. The more complete the refocusing, the larger the signal and more accurate the reconstruction of the spectrum. The original KBB scheme used pulses with constant amplitudes and a linear frequency sweep over the designated bandwidth. Performance can be further improved by shaping the pulses to have an adiabatic hyperbolic secant (HS) amplitude with the frequency swept according to a hyperbolic tangent \cite{silver_highly_1984,fu_evaluation_1996,park_spin-echo_2009,endeward_implementation_2023}. In the last decade, pulse shaping has become possible in EPR as arbitrary waveform generators (AWGs) became fast enough to generate shaped pulses that cover bandwidths larger than those obtainable by hard square pulses. For this, microwave pulse amplitude and phase are modulated with sub-ns timing resolution. Currently, AWGs with sampling rates of 1.25 GS/s or faster are in use.

However, HS pulses are not optimal. In practice, limited available power limits the maximum achievable pulse amplitude which in turn limits the frequency sweep rate and therefore puts a lower bound on the pulse time. However, relaxation can put an upper limit on pulse duration. Also, in some sequences, pulse lengths must be shorter than the evolution periods of the interactions of interest, for instance dipolar couplings in dipolar EPR. \cite{endeward_implementation_2023}. Thus, along with the constraint of limited power, constraining the pulse time can cause the performance of pulses to suffer.

In response, optimal control methods such as composite pulses \cite{counsell_analytical_1985,levitt_composite_1986}, adiabatic pulses \cite{baum_broadband_1985,garwood_return_2001}, optimal control theory (OCT) pulses using different numerical algorithms \cite{cano_adjustable_2002,skinner_application_2003,kobzar_exploring_2004,foroozandeh_improved_2019,haller2022sordor}, including gradient ascent pulse engineering (GRAPE) \cite{khaneja_optimal_2005,nielsen_optimal_2010,tosner_optimal_2009,borneman_bandwidth-limited_2012,goodwin_modified_2016,spindler_shaped_2012}, were developed to accomplish broadband excitation and uniform inversion across a given bandwidth. In practice, optimal shaped pulses are distorted by nonlinearities in the power amplifier and by the resonator transfer function, moving them away from the extremum in the optimization landscape. One can try to compensate for these distortions post-optimization, but the necessary compensation may not be possible while respecting constraints such as limited power at fixed pulse time. 

In this paper, we use an optimization method that includes a model of the full experimental transfer chain and resulting shape distortions while limiting both the available power along with the length of pulses. We investigate the use of variously parameterized broadband pulses for a Hahn echo sequence and the effect of varying the pulse length ratio on echo amplitude, refocusing time, and refocusing phase. Due to the freedom of the large parameter space and low number of constraints, we find the individual pulses ($\pi/2$ and $\pi$) act cooperatively as done previously with NMR COOP pulses \cite{braun2010cooperative,wei2014cooperative,asami2018ultrashort,kallies2018cooperative}. This shows that cooperatively performing pulses are still optimal under transmission distortions and in the presence of a power constraint while in limited pulse length time. Performances of individual pulses as well as of the entire pulse sequence are compared. Section II summarizes the transmitter model and the spin physics model, section III describes the pulse parameterizations used, section IV provides details about the optimization method, and section V discusses the results.

\begin{figure*}
\includegraphics{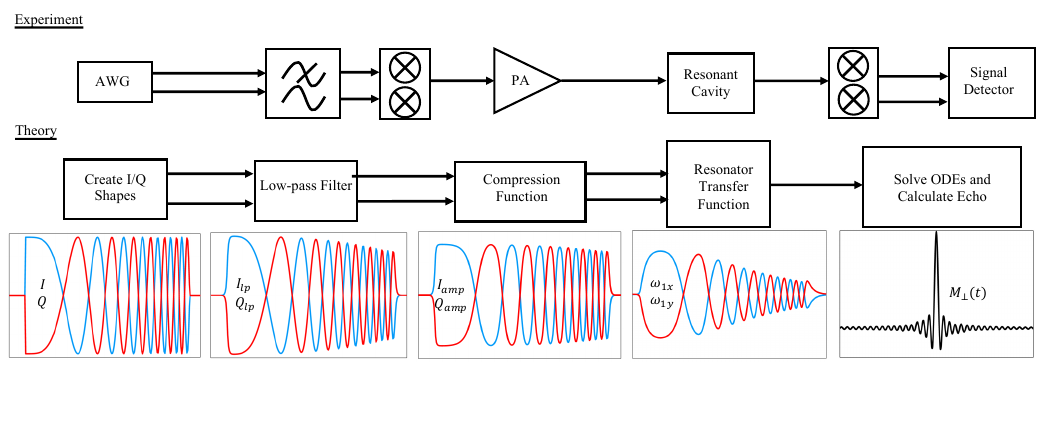}
\caption{A schematic overview of the experimental transmitter setup (top), its computational representation (middle) and exemplary signals (bottom). The input I/Q shapes are modeled and sent through a low-pass filter, amplifier compression function, and a resonator transfer function before they are used in the spin quantum dynamics calculation of the echo.}
\label{fig:blockdiagram}
\end{figure*}

\section{Model}\label{sec:model}

In order to model distortions that affect the pulse shapes, we closely follow the transmitter design of a typical EPR spectrometer. This is illustrated in Fig.~\ref{fig:blockdiagram}. First, AWG-generated in-phase and quadrature drive functions $I(t)$ and $Q(t)$ with amplitudes in the range $\left[-1, 1\right]$ are set up using different parameterizations described in detail in the next section. We model the limited output bandwidth of the AWG by applying a low-pass filter with transfer function
\begin{equation}\label{eq:lowpass}
H_\mathrm{lp}(\omega)
=
\frac{1}{1+\big(\omega/\varGamma\big)^2}.
\end{equation}
to $I$ and $Q$ via convolution, yielding
\begin{equation}\label{eq:lowpass distortion}
I_\mathrm{lp}(t) =
V_\mathrm{DAC} \mathcal{F}^{-1}\left[H_\mathrm{lp}(\omega) \cdot \mathcal{F}\left[I(t)\right]\right]
\end{equation}
where $\mathcal{F}$ represents the Fourier transform, and $V_\mathrm{DAC}$ represents the conversion factor from the digital-to-analog converter (DAC). $\varGamma$ represents the $3$-dB bandwidth of the filter. The $Q$ channel pulse shape is similarly distorted into $Q_\mathrm{lp}(t)$. The conversion factor between the dimensionless input and the voltage output is combined with other overall multiplicative factors at the end of the transmission chain model.

The IQ upconversion of the low-pass filtered $I_\mathrm{lp}(t)$ and $Q_\mathrm{lp}(t)$ to the carrier frequency $\omega_\mathrm{c}$ yields
\begin{equation}
V(t)
=
I_\mathrm{lp}(t)\cos(\omega_\mathrm{c}t)
\pm
Q_\mathrm{lp}(t)\sin(\omega_\mathrm{c}t)
\end{equation}
where the sign depends on whether the carrier reference for the Q channel mixer is phase shifted $+90$ or $-90$ degree relative to the carrier for the I channel mixer.

The upconverted signal can be rewritten in terms of an amplitude- and phase-modulated oscillation at the carrier frequency:
\begin{equation}
V(t) =
V_0(t)
\cos(\omega_\mathrm{c}t+\varphi(t))
\end{equation}
with the time-varying amplitude 
\begin{equation}
V_0(t) = \sqrt{I_\mathrm{lp}^2(t)+Q_\mathrm{lp}^2(t)}
\end{equation}
and phase
\begin{equation}
\varphi(t) = \mathrm{atan2}(Q_\mathrm{lp},I_\mathrm{lp}),
\end{equation}
where $\mathrm{atan2}$ is the two-argument arctangent.
The amplifier amplifies and additionally distorts this signal, leading to the appearance of higher harmonics in the amplifier output, $V_\mathrm{amp}(t)$. Assuming that the amplifier is memory-less, that the higher harmonics are rejected by the narrow-band transmission lines, and assuming separation of timescales, i.e. $V_0(t)$ and $\varphi(t)$ vary much more slowly than the carrier signal $\cos(\omega_\mathrm{c} t)$, the amplified signal is represented by 
\begin{equation}\label{Vamp function}
V_\mathrm{amp}(t)
=
G\!\left(V_0(t)
\right) V_0(t)
\cos(\omega_\mathrm{c}t+\varphi(t))
\end{equation}
where $G$ is the gain function, or in terms of $I_\mathrm{lp}$ and $Q_\mathrm{lp}$
\begin{equation}
V_\mathrm{amp}(t)
=
I_\mathrm{amp}(t)\cos(\omega_\mathrm{c}t)
+
Q_\mathrm{amp}(t)\sin(\omega_\mathrm{c}t)
\end{equation}
with%
\begin{equation}
I_\mathrm{amp}(t)
=
G(V_0(t))I_\mathrm{lp}(t)
\end{equation}
and a similar expression for $Q_\mathrm{amp}(t)$.

These equations only model amplitude-to-amplitude modulation (AM/AM) effects of the amplifier and neglect possible amplitude-to-phase modulation (AM/PM) effects. AM/PM effects would alter the term $\varphi(t)$ in Eq.~\eqref{Vamp function} by mixing the I and Q signals.

In order to carry out numerical optimizations we have to specify a gain function, and we will use
\begin{equation}\label{eq:compression}
G\!(V_0(t)) = g\frac{\tanh{(V_0(t)/V_\mathrm{sat})}}{V_0(t)/V_\mathrm{sat}},
\end{equation}
where $g$ is the small-signal gain factor and $V_\mathrm{sat}$ the input saturation amplitude. $V_\mathrm{sat}$ parameterizes the nonlinearity: for $V_0(t) \ll V_\mathrm{sat}$ the amplifier is in the linear regime, while for $V_0(t) \gg V_\mathrm{sat}$ the amplifier is saturated. We assume $V_\mathrm{sat}$ is constant over the amplifier bandwidth and that the amplifier bandwidth is wider than the signal bandwidth. In principle, we could use different nonlinear models such as Rapp, Saleh, or polynomial models \cite{10366299}, or a tabulated function.

Next, the amplified pulse is transmitted to the resonator. The resonator transfer function is well described by
\begin{equation}
H_{\mathrm{res}}(\omega)
=\frac{1}{
1+\mathrm{i}Q_\mathrm{L}
(\frac{\omega}{\omega_{\mathrm{res}}}
-
\frac{\omega_{\mathrm{res}}}{\omega})
},
\end{equation}
where $Q_\mathrm{L}$ is the loaded Q-value and $\omega_{\mathrm{res}}$ is the resonator frequency. This produces the following pulse shape at the sample inside the resonator,
\begin{equation}\label{drive shape}
B_1(t)
= 
C\operatorname{Re} \left(\mathcal{F}^{-1}\left[H_{\mathrm{res}}(\omega) \cdot \mathcal{F}\left[V_\mathrm{phasor}(t)\right]\right]\right),
\end{equation}
where
\begin{equation}\label{eq:phasor}
V_\mathrm{phasor} (t) =
\left(
I_\mathrm{amp}(t) + \mathrm{i} Q_\mathrm{amp}(t)
\right)
\mathrm{e}^{\mathrm{-i}\omega_\mathrm{c}t}
\end{equation}
$C$ is the resonator conversion factor. In the sample, spins with gyromagnetic ratio $\gamma$ experience the drive function
\begin{equation}
\gamma B_1(t) = \omega_1(t) \cos\phi(t)
\end{equation}
with the drive amplitude $\omega_1$ and the phase $\phi$. With this transmitter chain, the maximum drive amplitude that can be achieved with $|I|=|Q|=1$ and $\omega = \omega_\mathrm{res}$ is $\omega_{1,\mathrm{max}}=\gamma gCV_\mathrm{sat} \tanh({\sqrt{2}V_\mathrm{DAC}/V_\mathrm{sat}})$. 

Note that there is a subtle difference between imposing separate constraints on $I$ and $Q$ as we do here versus simply imposing an overall amplitude constraint. Constraining $I$ and $Q$ independently restricts the accessible domain in the I/Q plane to a square rather than a circle. Maximum power is only attained at the four corners of the square when $|I|=|Q|=1$, i.e., phases of $\pm 45^\circ, \pm 135^\circ$. Changing the phase of a maximum-power pulse will generally result in a power loss. However, phase cycling can still be carried out in $\pm 90^\circ$ increments since this domain is symmetric under $\pm 90^\circ$ rotations.

The laboratory frame Hamiltonian for a given spin packet with Larmor frequency $\omega_{\mathrm{res}}$ is
\begin{equation}\label{eq:Hlab}
H_{\mathrm{lab}}(\omega_{\mathrm{res}},t)=
\omega_{\mathrm{res}}S_z +
\omega_1(t) \cos{(\phi (t))}S_x,
\end{equation}
where $S_i$ ($i$ = $x$, $y$, $z$) are spin operators. For EPR experiments, $\omega_{\mathrm{res}}/2\pi$ is typically on the order of 10--100 GHz while $\omega_1/2\pi$ is on the order of tens of MHz, so the rotating-wave approximation is valid when moving to a frame rotating at the carrier frequency, $\omega_\mathrm{c}$. The rotating-frame Hamiltonian for a spin packet off-resonant with the carrier frequency by $\Delta \omega = \omega_{\mathrm{res}} - \omega_{\mathrm{c}}$ is
\begin{equation}\label{eq:Hrot}
H_{\mathrm{rot}}(\Delta\omega,t) =
\Delta \omega S_z + 
\omega_{1x}(t) S_x +
\omega_{1y}(t) S_y,
\end{equation}
where 
\begin{align}
\omega_\mathrm{1x} = \omega_{1}(t) \cos{(\phi (t) - \omega_\mathrm{c}t)},
\\
\omega_\mathrm{1y} = \omega_{1}(t) \sin{(\phi (t) - \omega_\mathrm{c}t)}.
\end{align}

In experiments, the signal passes back through the resonator and, after downconversion, a bandwidth-limited video amplifier. The resulting receiver-side distortions affects all echo signals equally and the optimal echo shape, obtained when all spin packets align on the same axis along the $xy$-plane, does not change. Incorporating receiver distortions does not affect the pulse optimization, as the optimal echo shape is formed when all spins are aligned, regardless of receiver distortions, and so we do not include it in the model.

We now consider an ensemble of spin packets, each governed by a Hamiltonian with its own $\omega_\mathrm{res}$, ranging over a spectral distribution of width $\delta$. We choose to place $\omega_{\mathrm{c}}$ in the center of this distribution, so that $\Delta\omega$ ranges from $-\delta/2$ to $+\delta/2$. To obtain the effect of a given pulse we must solve an ensemble of Schr\"{o}dinger equations for the propagators $U$,
\begin{equation}\label{eq:Schrodinger}
\mathrm{i}\dot{U}(\Delta\omega,t) = H_{\mathrm{rot}}(\Delta\omega,t)U(\Delta\omega,t),
\end{equation}
corresponding to the different resonant frequencies across the relevant spectral width.

With our mapping between the I/Q inputs and the driving functions in the spin Hamiltonian, $\omega_{1x}$ and $\omega_{1y}$, we can solve Eq.~\eqref{eq:Schrodinger} for both the $\pi/2$ and $\pi$ pulses for a representative ensemble of Larmor frequencies within the desired band. Using the initial condition of $U(t=0)=\mathbb{1}$, where $\mathbb{1}$ is the identity matrix, we can solve the ensemble of differential equations, giving us a corresponding evolution operator for each Larmor frequency. The evolution operators for the $\pi/2$ and $\pi$ pulse are denoted as $U_{\pi/2}^i$ and $U_{\pi}^i$, where $i$ indexes the Larmor frequencies. We combine these propagators with the free evolution periods in the rotating frame to obtain the total propagators
\begin{equation}
U_{\mathrm{tot}}^i(t) =
U_{\mathrm{free}}^i(t)
U_{\pi}^i
U_{\mathrm{free}}^i(\tau_1)
U_{\pi/2}^i\quad
\end{equation}
where $t = 0$ now denotes the end of the $\pi$ pulse and the beginning of the second free evolution period. The signal due to a particular spin packet is 
\begin{equation}\label{eq:spinpacketsignal}
M_{\perp,i}(t)
=
2\,\mathrm{tr}
\left(
U_{\mathrm{tot}}^i(t)
\rho_0
U_{\mathrm{tot}}^{i\dag}(t)
S_+
\right),
\end{equation}
where $\rho _0$ is the initial density matrix for each spin, $\rho_0 = \mathbb{1}/2 - S_z$. 

The total magnetization signal from all spins is given by averaging over all spin packets,
\begin{equation}\label{eq:total echo}
M_{\perp}(t) =
\frac{1}{N}\sum_{i=1}^N
M_{\perp,i}(t),
\end{equation}
where in our optimization the Larmor frequencies were sampled from a uniform distribution. Our goal is to determine optimal shapes of $I(t)$ and $Q(t)$ that maximize the total magnetization signal, $M_{\perp}(t)$. As a check on the spin physics, all of the pulses generated were also tested and confirmed using the {\sc matlab} package Easyspin, an open-source software that allows for the simulation and analysis of EPR spectra \cite{stoll_easyspin_2006,stoll_general_2009}. The optimized pulse shapes and code are publicly available \cite{data}.

\begin{figure}
\includegraphics{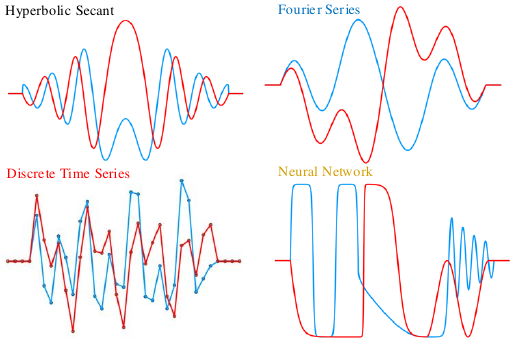}
\caption{Four different parameterizations of a pulse shape (blue: in-phase, red: out of phase).}
\label{fig:Fig2}
\end{figure}

\section{Parameterizations}\label{sec:parameterizations}

We start by considering the original KBB broadband pulse sequence using generalized HS shapes for the $\pi/2$ and $\pi$ pulses. KBB is not the only broadband refocusing sequence we could use. For instance, the CHORUS sequence \cite{power2016increasing,power2016very} uses linear swept pulses with effectively rectangular amplitude profiles that have more integrated power than hyperbolic secant pulses for a fixed $B_1$ amplitude and sequence time, and features improved robustness to $B_1$ field inhomogeneity \cite{foroozandeh_improved_2019,verstraete2021chirped}. However, we consider a strongly power-constrained regime with limited amplitude and pulse lengths where $B_1$ field inhomogeneity is a secondary concern. In this regime, HS pulses are preferable because they require less power off resonance, where it is costly to compensate for the profile of the resonator transfer function. Therefore, in this work we will use the KBB sequence with HS pulses as a baseline against which to compare other shaped pulses.

Then, while retaining the general bipartite structure of a $\pi/2$ pulse followed by a $\pi$ pulse, to introduce more shape flexibility, we consider three models with significantly more parameters: a nonlinear amplitude-limited Fourier series (FS), a discrete-time series (DT), and a neural network (NN). We do not constrain the pulse flip angles to be $\pi/2$ and $\pi$, but we will still refer to them by those labels in continuity with the KBB design and in anticipation that the optimization process will indeed drive them to be such.

\subsection{Generalized HS pulses}\label{subsec:sech}

For representing generalized HS pulses, we use excitation functions of the form \cite{tannus1996improved,doll_wideband_2017}
\begin{align}\label{eq:HS}
\omega_1(t) &=
A
\operatorname{sech}
\left(
2^{n-1}\beta\left|t/T\right|^n
\right)
\\
\dot\phi_{\mathrm{HS}}(t) &= \Delta\omega_\mathrm{BW}
\left(
\frac{
\int_{-T/2}^t\operatorname{sech}^2\!\left(2^{n-1}\beta\left|\tau/T\right|^{n}\right)\mathrm{d}\tau
}{
\int_{-T/2}^{T/2}\operatorname{sech}^2\left(2^{n-1}\beta\left|\tau/T\right|^{n}\right)\mathrm{d}\tau} - \frac{1}{2}
\right),
\end{align}
where $-T/2\le t\le T/2$, and $\omega_1(t)$ and $\dot\phi_\mathrm{HS}(t)$ are the amplitude and instantaneous driving frequency at time $t$ in the rotating frame. Although $n$ is usually considered to be a positive integer ($n=1$ in the original KBB scheme), we extend the definition to include non-integer $n$ by using the absolute value of the time. Similarly, while $\Delta\omega_\mathrm{BW}$ is typically set to the desired excitation bandwidth, we allow this to be a free parameter as well.

These functions are converted to $\omega_{1x}$ and $\omega_{1y}$ used in Eq.~\eqref{eq:Hrot} by using
\begin{align}\label{eq:I/Q conversion}
\omega_{1x} &= \omega_1(t) \cos(\phi_\mathrm{HS}(t)),
\\
\omega_{1y} &= \omega_1(t) \sin(\phi_\mathrm{HS}(t)),
\end{align}
with the phase $\phi_\mathrm{HS}(t)$ given by
\begin{equation}\label{eq:cumulative phase}
\phi_\mathrm{HS}(t) = \int_{-T/2}^{t} \dot\phi(\tau) d\tau.
\end{equation} 
The $\pi/2$ and $\pi$ pulses of this form are chosen such that, following KBB, the two pulse durations have the ratio $T_{\pi/2}/T_{\pi} = 2$. The values of $\beta, n$, and $\Delta\omega_{\mathrm{BW}}$, along with $\omega_{1x}$ and $\omega_{1y}$ for each of the two pulses are free parameters, and we optimize them for refocusing spins across a bandwidth of $\delta$. We do not require the two pulses to produce $\pi/2$ and $\pi$ rotations for any spin packet, but only restrict $A \leq \omega_{1,\mathrm{max}}$.

To benefit from the intuitive behavior of HS pulses, we have to directly parameterize the pulse at the \emph{output} of the transmitter chain of Fig.~\ref{fig:blockdiagram}, then invert the transfer functions in order to obtain the I/Q inputs that should be programmed into the AWG. In contrast, the other parameterizations considered below are for the \emph{input} I/Q pulse shapes themselves and do not require pre-compensation. Also note that the HS pulses satisfy $I(t)^2 + Q(t)^2 \leq 1$, lying within the inscribing circle of the square domain $(|I(t)|\leq1, |Q(t)|\leq1)$ available to the other parameterizations on the I/Q plane, so the HS parameterization is clearly at a disadvantage to begin with because it cannot access as much power as a parameterization that allows, for example, $I=Q=1$. However, in order to focus on the less obvious differences between parameterizations, we have allowed the HS pulses to access the \emph{circumscribing} circle, $I(t)^2 + Q(t)^2 \leq 2$.

\subsection{Nonlinear Amplitude-Limited Fourier series}\label{subsec:fourier}

The second pulse shape model we consider consists of a nonlinear amplitude-limited Fourier series (FS) for the two drive functions
\begin{align}\label{eq:Fourier representation}
I(t) &= \tanh\!\left[{\sum_{n=1}^N a_{\mathrm{I},n} \cos{\left(\pi n \frac{t}{T}\right)}}\right]
\\
Q(t) &= \tanh\!\left[{\sum_{n=1}^{N} a_{\mathrm{Q},n} \cos{\left(\pi n\frac{t}{T}\right)}}\right]
\end{align}
for each pulse. Here again $-T/2\le t\le T/2$, and $a_{\mathrm{I},n}$ and $a_{\mathrm{Q},n}$ are real-valued coefficients.

The enclosing $\tanh$ function limits $I$ and $Q$ to values between $-1$ and $1$ by construction. The reasoning behind imposing the amplitude limit using $\tanh$, as opposed to scaling the coefficients, is that the maximum of the Fourier series signal can only be determined using a numerical search, which renders the cost function itself non-differentiable. Due to the nonlinearity of $\tanh$, the bandwidths of $I$ and $Q$ are not straightforwardly related to the frequencies included in the cosine series, and the modeled shapes are nonlinear and compressed compared to a standard cosine series. We choose $N$ large enough to cover the desired bandwidth, $\pi N/T \approx \delta$. Including higher-order terms in the series does not improve performance, as those terms are severely attenuated by the resonator in the relevant case where $\delta$ is comparable to resonator bandwidth $\omega_{\mathrm{res}}/Q_\mathrm{L}$. We perform an unconstrained optimization over $(\boldsymbol{a}_{\mathrm{I}}, \boldsymbol{a}_{\mathrm{Q}})$ for each pulse.

\subsection{Discrete-time series}\label{subsec:discreteT}

The third pulse shape model we consider consists of discrete-time series (DT) for $I$ and $Q$ for each pulse
\begin{equation}\label{eq:pointwise representation}
\boldsymbol{I}
=
\tanh(\boldsymbol{a}_\mathrm{I}),
\qquad
\boldsymbol{Q}
=
\tanh(\boldsymbol{a}_\mathrm{Q}),
\end{equation}
where $\boldsymbol{I}$ and $\boldsymbol{Q}$ are vectors with elements $I_i = I(i\Delta t)$ and $Q_i = Q(i\Delta t)$ for $i = -N$ to $N$. The time increment $\Delta t$ is chosen to be on the order of sub-ns, in accordance with the sampling rates of modern AWGs \cite{brown_broadband_2008,tseitlin_digital_2011}. Just as before, we use an element-wise tanh as an enclosing function to constrain the values to between $-1$ and 1. The $\pi/2$ and $\pi$ pulses are parameterized by separate $(\boldsymbol{a}_\mathrm{I},\boldsymbol{a}_\mathrm{Q})$.

\subsection{Neural network}\label{subsec:nn}

\begin{figure}
\includegraphics{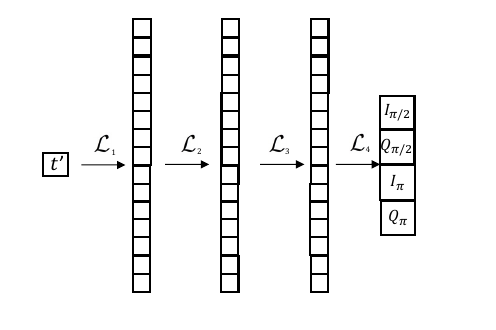}
\caption{A representation of the neural network utilized where the input layer is a single node consisting of the time $t'$, $3$ hidden layers of $16$ nodes each, and the output layer of the $4$ drive signal amplitudes at time $t'$.}
\label{fig:Fig3}
\end{figure}

Finally, we follow \cite{gungordu_robust_2022} in creating a deep neural network (NN) model to represent the pulse shapes. The model, represented in Fig.~\ref{fig:Fig3}, is
\begin{equation}\label{eq:neural network representation}
\begin{pmatrix}
I_{\pi/2}(t'T_{\pi/2})\\
Q_{\pi/2}(t'T_{\pi/2})\\
I_{\pi}(t'T_{\pi})\\
Q_{\pi}(t'T_{\pi})
\end{pmatrix}
=
\mathcal{L}_4 \circ \mathcal{L}_3 \circ \mathcal{L}_2 \circ \mathcal{L}_1(t'),
\end{equation}
where each layer $\mathcal{L}_i$ takes a $d_{i}$-dimensional input vector and maps it to a $d_{i+1}$-dimensional output vector according to the function
\begin{equation}
\mathcal{L}_i(\boldsymbol{x})
=
\tanh (\mathcal{W}_i\boldsymbol{x}+\boldsymbol{b}_i)
\end{equation}
with a $d_{i+1} \times d_i$ weight matrix $\mathcal{W}_i$, a $d_i$-dimensional bias vector $\boldsymbol{b}_i$, and tanh as an element-wise activation function that ensures the outputs are confined to between $-1$ and $1$. The first layer contains only a single node (i.e., $d_1 = 1$) and the input to it is the dimensionless time value $t' \in \left[-1/2,1/2\right]$. Layers 2, 3, and 4 have 16 nodes, and the final layer contains four nodes that output the values for the four functions $I_{\pi/2}(t' T_{\pi/2})$, $Q_{\pi/2}(t' T_{\pi/2})$, $I_{\pi}(t' T_{\pi})$, and $Q_{\pi}(t' T_{\pi})$. We optimize the model parameters $\boldsymbol{b}_i$ and $\mathcal{W}_i$, 644 in total. In this NN model, all four pulse shapes are controlled by the same set of parameters, which enables the model to represent possible correlations between the two pulses.

\section{Optimization}\label{sec:cost function}

In order to efficiently represent the distortion chain in our cost function, we utilize fast Fourier transforms (FFT), sampling the continuous shapes of the HS, FS and NN models with the same time step, $\Delta t$, as in in the DT model. To obtain the total echo amplitude, we use a numerical solver to solve the ensemble of Schr\"{o}dinger equations \eqref{eq:Schrodinger} for $250$ frequencies spaced equidistantly across the desired band for a large enough time range to encompass any refocusing point.

For each model, we optimize over the full vector of parameters $\boldsymbol{p}$ to maximize the echo amplitude irrespective of echo phase. The objective function is
\begin{equation}\label{eq:cost function value}
    \mathcal{J}(\boldsymbol{p}) =
    {\underset{t}{\mathrm{max}}}
    |M_{\perp}(t)|.
\end{equation}
In a KBB pulse sequence, the echo occurs at time $t=T_{\pi} + \tau_1$ after the end of the $\pi$ pulse. However, for the other pulse shape models, the echo can occur at earlier or later times. 

The objective function used for the optimization of the hyperbolic secant functions is slightly different. As mentioned earlier, because the HS pulse shapes are the \emph{output} I/Q pulse shapes, we need to ensure that the corresponding \emph{input} I/Q pulse shapes respect the power constraint. For instance, if the HS pulse had an amplitude of $\omega_{1,\mathrm{max}}$ while the instantaneous driving frequency is off-resonant with the resonator, the input pulse required to compensate for the resonator transfer function would exceed $\omega_{1,\mathrm{max}}$. Thus, we include an extra term in the objective function to penalize the pulse for exceeding the power limit at any of the sampling points, 
\begin{equation}\label{eq:HS cost function value}
    \mathcal{J}(\boldsymbol{p})_{\mathrm{HS}} =
    \mathcal{J}(\boldsymbol{p}) -
    \sum_{i=0}^{T/\Delta t}
    \mathrm{max}\left(0,\omega_{\mathrm{in}}^i -\omega_{1,\mathrm{max}}\right),
\end{equation}
where $\omega_{\mathrm{in}}^i$ is the amplitude at time $i\Delta t - T/2$ of the pulse when compensating for the resonator transfer function,
\begin{equation}\label{input amplitude calculation}
\omega_{\mathrm{in}}(t) = \mathcal{F}^{-1}\left[\mathcal{F}\left[\omega_1(t)\right]/H_{\mathrm{res}}(\omega)\right],
\end{equation}
with $\omega_1(t)$ as given in Eq.~(\ref{eq:HS}). We leave out the factor of $\Delta t$ in the second term in Eq.~\eqref{eq:HS cost function value} in order to more heavily weight this term in the cost function to enforce the amplitude constraint. In practice, the optimization of $\mathcal{J}(\boldsymbol{p})_{\mathrm{HS}}$ leads to the second term being zero and the optimal value of the objective function is the same as the echo amplitude.

For the FS parameterization, we use $24$ terms per pulse shape, totaling $96$ free parameters for the sequence. For the DT parameterization, we use a time step of $\Delta t = 0.625$ ns, resulting in $384$ parameters for $T_{\pi/2}+T_\pi= $ 120 ns. The NN parameterization uses $644$ free parameters, as described in Sec.~\ref{subsec:nn}. We used the Julia package DiffEqFlux.jl \cite{rackauckas_universal_2021} to form the NN parameterization and the BFGS optimizer from the Zygote.jl and Optim.jl packages for optimizing the various pulse shape parameterizations \cite{innes_dont_2019}. The limiting factor in computational cost is the numerical solution of Schr\"{o}dinger's equation for each of the $250$ Larmor frequencies for both pulses at each optimization step. We solved these in parallel on cluster computing resources with the Bogacki--Shampine method (BS5) as implemented in the DifferentialEquations.jl package. All optimizations were terminated upon the condition that the difference between the current objective function value and the objective function value $20$ steps previous was less than $10^{-4}$. All three large parameterizations took around $500$-$1000$ optimization steps to converge, whereas the 5-parameter hyperbolic secant model only took tens of steps. On average, these optimizations took around $2$-$3$ hours to complete running with $36$ cores in parallel. However, this substantial optimization time is not a problem, as it is a one-time computational cost for a given spectrometer setup. As long as the distortion chain has been properly characterized, the computed pulses should work without (or with minimal) spectrometer-based feedback optimization \cite{goodwin2018feedback}. The robustness of these pulses to mischaracterization of the distortion chain elements and $B_1$ field inhomogeneity is discussed in Section \ref{sec:results}.
\begin{figure}
\includegraphics{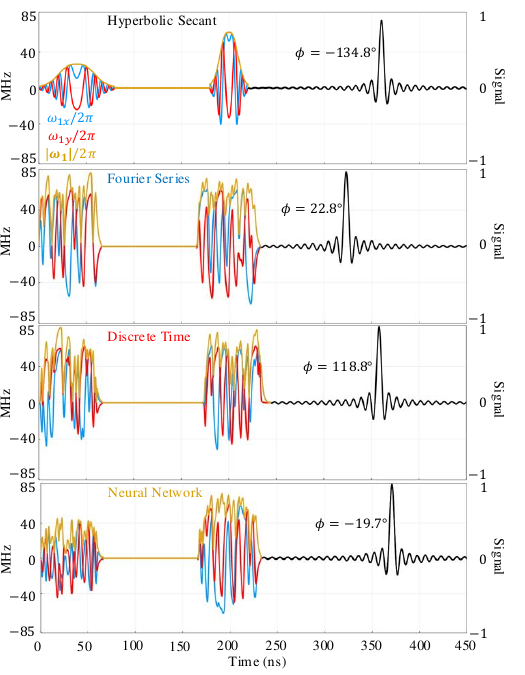}
\caption{The optimized pulse shapes as seen by the spins after passing through the chain of transfer functions shown in Fig.~\ref{fig:blockdiagram}, along with the formed echoes, the real part of the magnetization after rephasing the echo such that the maximum is completely real. The optimized HS $\pi/2$ and $\pi$ pulse lengths are $80$ and $40$ ns long respectively while the FS, DT, and NN have $\pi/2$ and $\pi$ pulse lengths of $60$ ns each. The phase, $\phi$, of each echo is also reported.}
\label{fig:Fig4}
\end{figure}

\begin{figure}
\includegraphics{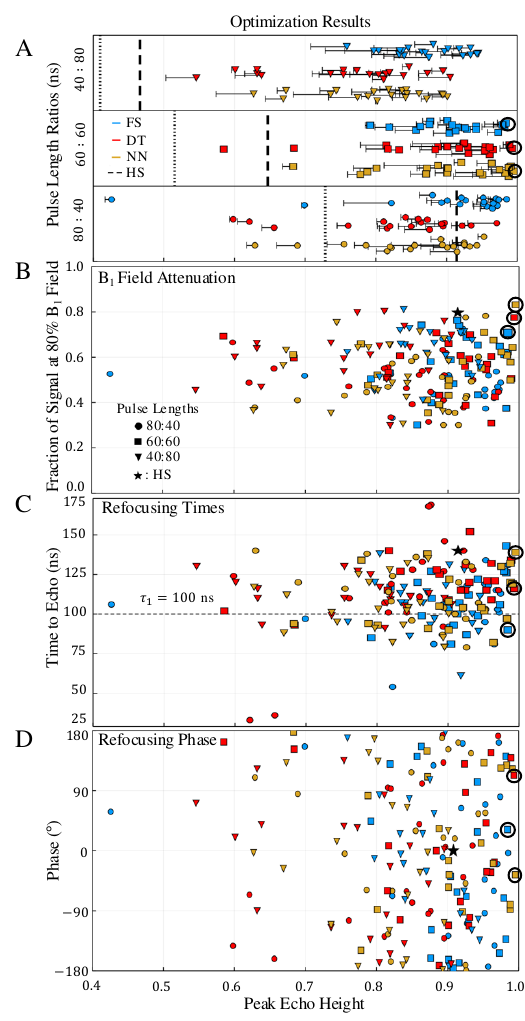}
\caption{Results of 20 optimizations each for the three large parameterizations of Fourier series (FS), discrete time (DT), and neural network (NN) with $80/40$ ns, $60/60$ ns, and $40/80$ ns pulse lengths from random starting points. The top panel shows the echo amplitudes for the different pulse types, where the dashed lines are the echo amplitudes of the best performing hyperbolic secant (HS) pulses for each respective pulse length ratio. The $3$ circled points represent the $3$ best performing pulse shapes for each parameterization that are plotted in Fig.~\ref{fig:Fig4}. The error bars and dotted lines show the effect of altering the amplifier compression function as described in the text. The bottom three panels show the relative echo amplitude reduction for the case where $B_1$ field is reduced to $80\%$ of its original value, the time between the end of the $\pi$ pulse and the echo maximum, and the phase with which the spins are refocusing, all versus the echo amplitudes.}
\label{fig:Fig5}
\end{figure}

\section{Results}\label{sec:results}

We performed the optimizations using the parameter values shown in Table \ref{Table1}, representative of a Q-band EPR spectrometer with a nitroxide sample. The carrier frequency of $33.65$ GHz corresponds to a static magnetic field strength of about $1.2$ T. The value of $\varGamma$ corresponds to a $3$-dB bandwidth of $450$ MHz for the AWG. The loaded Q-value corresponds to a resonator $3$-dB bandwidth of $168$ MHz. The maximum power limit of $\omega_{1,\mathrm{max}}=84$ MHz corresponds to an oscillatory magnetic field strength of about $3$ mT. The choice of a 120 ns total pulse time was made in order to examine a case where the power constraint starts to deteriorate the performance of the hyperbolic secant pulse. The delay time, $\tau_1$, was chosen to be $100$ ns. While this value may affect the path the optimizer takes through the optimization landscape, pulse shapes optimized with one value of $\tau_1$ produce the same echo amplitude for a different value of $\tau_1$. The nonlinear phase dispersion of the $\pi/2$ pulse will still be cancelled by that of the $\pi$ pulse, and changing the delay time between them only changes the linear part of the phase dispersion, thus changing the echo time and phase but not the echo amplitude. 

The optimization and the resulting analysis does not consider any spins to be coupled. Chirped pulses in particular have been shown to create unwanted artifacts compared to rectangular pulses in situations with coupled electron and nuclear spins \cite{doll2016epr,pribitzer2016transverse}. A separate analysis can be performed to determine the effect that these optimized pulse shapes have on multi-dimensional spectra. 

For each of the parameterizations, we also considered three different ratios $T_{\pi/2}/T_\pi$, 2:1, 1:1, and 1:2, keeping the total pulse time fixed at 120 ns. The KBB sequence requires a 2:1 ratio, i.e. 80 and 40 ns for the $\pi/2$ and $\pi$ pulses, respectively. However, a $40$ ns $\pi$ pulse requires more power to adiabatically flip spins than the available limit, so many of the spins are under-rotated and the performance begins to suffer. Decreasing the time of the $\pi/2$ pulse while increasing the time of the $\pi$ pulse will alleviate this issue, but it will also cause the phase refocusing aspect of the KBB pulse sequence to suffer \cite{spindler_perspectives_2017, bowen2018orientation}. Here, we have optimized the HS pulses for 1:1 and 1:2 pulse length ratios to demonstrate this tradeoff inherent to the KBB design. 

The best HS result overall was for a 2:1 pulse length ratio, with $A_{\pi/2} / 2\pi = 23.42$ MHz and $A_{\pi} / 2\pi = 61.66$ MHz for the amplitudes of the $\pi/2$ and $\pi$ pulses, $\beta = 7.09$, $n = 1.57$, and $\Delta\omega_{\mathrm{BW}} / 2\pi = 138.51$ MHz, which produces an echo amplitude of $0.9070$ (see Fig.~\ref{fig:Fig4} top). To quantify the effect of artificially allowing the HS to exceed the power limit as discussed in Sec.~\ref{sec:cost function}, we also optimized strictly within the inscribing circle of the allowed domain in the I/Q plane, $I(t)^2 + Q(t)^2 \leq 1$. This more restricted optimization resulted in an echo amplitude of $0.8621$ for our example parameters. Of course, the total extra power gained from accessing the corners of the square domain diminishes as the value $V_\mathrm{sat}$ decreases, since the amplifier saturates more readily. Having quantified this effect, below we set it aside and focus on the performance of other parameterizations compared to the HS with equivalent maximum power whose echo amplitude is $0.9070$. As an alternative point of comparison, the time length required for an HS pulse sequence to perform as well as the other optimized pulse shapes is $180$ ns, $50\%$ longer than the other pulses.

For those other pulse parameterizations, all three pulse length ratios do allow for both full rotation across the bandwidth and refocusing, with a 1:1 pulse length ratio performing the best. The maximized echo amplitudes are, in increasing order, $0.9852$ for the FS, $0.9928$ for the DT, and $0.9951$ for the NN. These pulses plotted in Fig.~\ref{fig:Fig4} are the drive functions seen by the spins, i.e. after passing through the low-pass filter, the amplifier, and the resonator transfer function (thus the pulses are several ns longer than the nominal $60$ ns due to the finite ring-down time). The pulses show irregular shapes; there are no apparent interpretable features. Clearly, the FS, DT, and NN pulses offer performance gains compared to the HS for this scenario of short pulse time and limited power. Among these three parameterizations, though, there is not one that stands out as particularly advantageous. As mentioned in Sec.~\ref{sec:parameterizations}, the reasoning behind using the NN was to efficiently represent possible correlations between the two pulses. However, both the FS and DT show equivalent cooperativity in compensating for phase accumulation between the $\pi/2$ and $\pi$ shapes, and the NN parameterization offered no extra advantage in that regard. The pulse shapes before passing through the distortion chain are plotted in Fig.~S1 of the Supplementary Material \cite{sup}.

All three parameterizations are equivalently efficient computationally, taking around the same number of optimization steps. The reason for the wide range of performance with many equivalent maxima is that the landscape contains many local maxima. Any large parameterization flexible enough to access a large area of that landscape in an unbiased way will lead to many maxima of varying quality. The outcome of a particular optimization depends on the initial point, so it is clearly useful to try many random initial seeds as we have done.

Figure \ref{fig:Fig5}A shows the optimized echo amplitudes for the three parameterizations and three pulse length ratios. For each combination of shape and duration, the results of 20 separate optimization runs with random starting points are shown. The plots show that for any ratio the best FS, DT, and NN pulse shapes all outperform even the best $2:1$ HS pulse (indicated by vertical dashed lines). For the FS, DT, and NN parameterizations, the best performing pulses have $1:1$ pulse length ratios.

A similar observation regarding pulse length ratios was made by Kallies and Glaser \cite{kallies2018cooperative}, where they found an optimal ratio of $1:1.3$ for their set of parameters. They used a different set of constraints, e.g., $\omega_{1,\mathrm{max}}/\delta = 0.2$ and $\omega_{1,\mathrm{max}}\cdot(T_{\pi/2}+T_\pi) = 6$, compared to our $\omega_{1,\mathrm{max}}/\delta = 0.35$ and $\omega_{1,\mathrm{max}}\cdot(T_{\pi/2}+T_\pi) = 10$. However, the two scenarios are more similar than these numbers would suggest because Ref.~\cite{kallies2018cooperative} did not account for the effect of a resonator and their pulses allowed a maximum pulse amplitude that is independent of the instantaneous driving frequency. Scaling their pulses to our desired bandwidth and compensating for low-pass filter and resonator, the required pulse amplitude is about twice as large in order to recover their pulse design when driving near the edges of the bandwidth. So, accounting for the transfer chain effectively makes their constraints to be $\omega_{1,\mathrm{max}}/\delta \approx 0.4$ and $\omega_{1,\mathrm{max}}\cdot(T_{\pi/2}+T_\pi) \approx 12$, similar to ours. 

The error bars in Fig.~\ref{fig:Fig5}A show the effect on the echo amplitude of reducing the amplifier compression $V_\mathrm{sat}$ by 60\%, i.e.~increasing amplifier compression without changing maximum output power. The FS, DT, and NN pulses are all relatively robust against this. This is because many of the numerically shaped pulses use as much power as possible, so only the maximum power of the amplifier matters rather than the shape of the compression function at intermediate power. This is not entirely clear from the plots of the output pulse shapes in Fig.~\ref{fig:Fig4}, but the input pulse shapes of $I$ and $Q$ plotted in Fig.~S1 of the Supplementary Material \cite{sup} are typically toggling between their maximum values of $-1$ and $1$, and the values plotted are only diminished from maximum amplitude due to the effects of the low-pass filter and resonator. Thus, changing the general shape of the amplifying function does not affect the pulse shapes much. The HS pulse shapes, on the other hand, do use the full range of the amplifier, and the dotted lines show how their performance is substantially diminished under the same $V_\mathrm{sat}$ reduction.

In the $B_1$ field attenuation plot in Fig.~\ref{fig:Fig5}B, we plot the fractional reduction of the echo amplitude when the $B_1$ field is reduced to $80\%$. This probes how robust the echo amplitude is to $B_1$ field inhomogeneity. The best HS pulse is one of the more robust pulses against $B_1$ field inhomogeneity, since it is constructed as an adiabatic frequency sweep. Other pulse types include equivalently robust pulses, but generally there is no correlation between echo amplitude and robustness to inhomogeneity because the pulses were not optimized for robustness. In Fig.~S3 of the Supplementary Material \cite{sup}, we also plot echo amplitude versus resonator quality factor and versus $B_1$ field inhomogeneity. 

In the refocusing times plot of Fig.~\ref{fig:Fig5}C, we plot the time between the end of the $\pi$ pulse and the peak of the echo signal for $\tau_1 = 100$ ns. Most refocusing times fall in the range between 80 and 140 ns. Similarly to the results of Ref.~\cite{kallies2018cooperative}, the middle of this optimal range is slightly longer than the waiting time $\tau_1$. There is no correlation of echo refocusing time with pulse parameterization or pulse length ratio.

The refocusing phase plot in Fig.~\ref{fig:Fig5}D shows no evidence for a predominant echo phase. This is surprising, since one could imagine echo phase preferences in the presence of a the $|I|\leq1$, $|Q|\leq1$ power constraint, where the maximum drive amplitude can only be achieved with driving phases $\pm 45^\circ,\pm 135^\circ$.

This ensemble of optimizations from different random seeds shows that there is no unique pulse shape that is optimal. Of course, there is always a trivial degeneracy of rotating both $I$ and $Q$ by $\pm 90^\circ, \pm 180^\circ$ resulting only in a change of echo phase by the same angle (the degeneracy is four-fold rather than continuous due to the power constraint being a square rather than a circle in the I/Q plane), retaining the same refocusing time and echo amplitude. But beyond that, Fig.~\ref{fig:Fig5} shows that there are many pulse shapes with equivalent echo amplitudes that are not simply related by a phase transformation, as is clear from the different echo refocusing times. The fact that the landscape harbors many comparable maxima suggests that there remains a significant amount of flexibility in the parameterized pulses that could be used to satisfy additional constraints, such as favoring a particular refocusing time or phase (as in \cite{kallies2018cooperative}), robustness to $B_1$ inhomogeneity, echo phase independence from $\tau_1$, or some other desirable property.

\begin{table}
\begin{tabular}{|c|c|} 
\hline
Quantity & Value \\
\hline
$\omega_\mathrm{c}/2\pi$ & $33.65$ GHz \\ 
$\delta/2\pi$ & $240$ MHz \\ 
$\varGamma/2\pi$ & $450$ MHz \\ 
$V_\mathrm{sat}/V_\mathrm{DAC}$ & $1.131$ \\
$Q_\mathrm{L}$ & $200$ \\
$\tau_1$ & $100$ ns \\
$\omega_\mathrm{res}/2\pi$ & $33.65$ GHz \\
$T_{\pi/2} + T_{\pi}$ & $120$ ns \\
$\omega_{1,\mathrm{max}}/2\pi$ & $84$ MHz \\
\hline
\end{tabular}
\caption{\label{Table1} A list of all the values used in the optimizations.}
\end{table}

\begin{figure}
\includegraphics{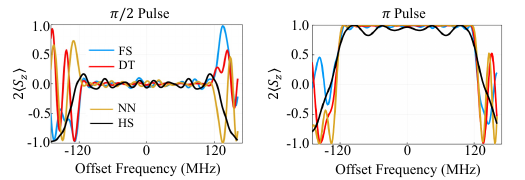}
\caption{The excitation profiles of the optimized pulses for the $\pi/2$ pulse (left), and the $\pi$ pulse (right), calculated from Eq. \eqref{eq:Excitation Profiles}.}
\label{fig:Fig6}
\end{figure}

In Fig.~\ref{fig:Fig6} we examine the action of the individual pulses by plotting the $z$-projection of a spin packet after it is rotated from the ground state by the shaped $\pi/2$ pulse,
\begin{equation}\label{eq:Excitation Profiles}
2\langle S_z (\omega) \rangle_{\pi/2}
=
2\,\mathrm{tr}(U_{\pi/2} \rho_0 U_{\pi/2}^{ \dag}S_z)
,
\end{equation}
where $\rho_0$ is the ground state density matrix, and similarly for the $\pi$ pulse. We plot over a frequency range extending slightly outside the $\delta$ bandwidth. 
Recall that we have not constrained the first and second pulse to be a $\pi/2$ and a $\pi$ pulse. Yet, Fig.~\ref{fig:Fig6} shows that inside the desired band the optimization always produces nearly perfect $\pi/2$ and $\pi$ pulses, suggesting this basic structure is optimal for refocusing. This is not an artifact of a particular initialization of the parameters -- we initialize randomly and the initial pulse shapes are not $\pi/2$ or $\pi$ rotations. We also see that for the FS, DT, and NN parameterizations, the effect of the pulses on spins outside the bandwidth varies wildly with frequency compared to the HS pulses which do not excite these spins.

Figure \ref{fig:Fig6} strongly suggests that the improvement of the optimal FS, DT, and NN pulses compared to the HS pulse comes from i) improved performance near the band edges and ii) improved intra-band $\pi$ rotation performance in the presence of the power constraint.
For a more comprehensive visualization of the spin dynamics, in Fig.~S2 of the Supplementary Material \cite{sup} we plot the total magnetization in $x$, $y$, and $z$ during each of the optimized pulse sequences.

\begin{figure}
\includegraphics{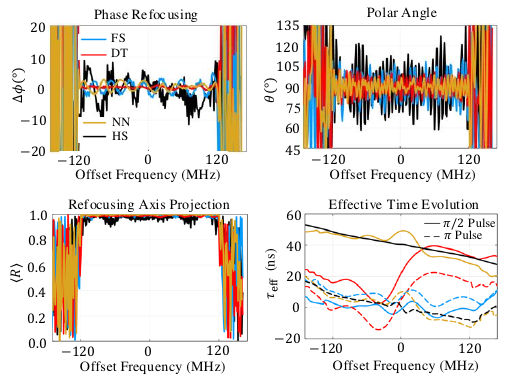}
\caption{Behavior of spin packets as a function of offset frequency. Phase difference $\Delta \phi$ relative to average phase at refocusing time (top left), polar angle $\theta$ at refocusing time (top right), projection onto the refocusing axis in the $xy$-plane at refocusing time (bottom left), and effective time evolution for each of the $\pi/2$ and $\pi$ pulses (bottom right) calculated from Eqs.\eqref{eq:phase dispersion}-\eqref{eq:Refocusing Projection} and \eqref{eq:Effective Time Evolutions} respectively.}
\label{fig:Fig7}
\end{figure}

In Fig.~\ref{fig:Fig7} we further characterize the action of the sequence as a whole on any given spin packet. The top left panel in Fig.~\ref{fig:Fig7} is a plot of the phase dispersion at the refocusing time. This is calculated by first computing the individual spin packet phases as a function of offset frequency,
\begin{equation}\label{eq:individual phase}
    \phi(\omega) = \arg \left( \mathrm{tr} (\rho_{\mathrm{refocus}}S_+^{\dag})\right),
\end{equation}
where
\begin{equation}
    \rho_{\mathrm{refocus}} = U_\mathrm{tot}(t_{\mathrm{refocus}})\rho_0
    U_\mathrm{tot}^{\dag} (t_{\mathrm{refocus}}),
\end{equation}
and $t_{\mathrm{refocus}}$ denotes the time at which the peak of the echo occurs. The average phase for spin packets within the band, which corresponds to the phase of the echo is
\begin{equation}\label{eq:average phase}
    \phi_{\mathrm{avg}} = \frac{1}{\delta} \int_{-\delta/2}^{\delta/2} \phi(\omega') \mathrm{d}\omega',
\end{equation}
and the phase dispersion plotted is
\begin{equation}\label{eq:phase dispersion}
    \Delta \phi(\omega) = \phi(\omega) - \phi_{\mathrm{avg}}.
\end{equation}
In other words, this is the azimuthal angle between the spin packet and the refocusing axis. 
The top right panel in Fig.~\ref{fig:Fig7} is a plot of the polar angle each spin packet forms with the $z$-axis at the refocusing time,
\begin{equation}\label{eq:amplitude refocusing}
    \theta(\omega) = \arccos \left( 2\,\mathrm{tr}(\rho_{\mathrm{refocus}}S_z)\right).
\end{equation}

From these top two panels we see that the phase dispersion and the polar angle with the FS, DT, and NN pulses are closer to ideal than with the HS pulse. The highly oscillatory polar angle compared to individual polar angles obtained in Fig.~\ref{fig:Fig6} is because the effect of a $\pi$ pulse on a spin packet depends upon the phase of the spin packet, which varies rapidly as a function of offset frequency due to the waiting time between the pulses.

In the bottom left panel of Fig.~\ref{fig:Fig7} we plot the projection of each spin packet onto the refocusing axis in the $xy$-plane,
\begin{equation}\label{eq:Refocusing Projection}
    \langle R \rangle = 2\,\mathrm{tr}\left(\rho_{\mathrm{refocus}}\left(\cos(\phi_{\mathrm{avg}}) S_x + \sin(\phi_{\mathrm{avg}}) S_y\right)\right).
\end{equation}
The FS, DT, and NN pulses are clearly more consistent in cooperatively producing an echo across the bandwidth.

In the bottom right panel of Fig.~\ref{fig:Fig7} we explore the cooperativity of the shaped $\pi/2$ and $\pi$ rotations using the effective evolution time defined in Ref.~\cite{kallies2018cooperative}. Consider the density matrix as a function of offset frequency at three intermediate times: immediately after the $\pi/2$ pulse, immediately before the $\pi$ pulse, and immediately after the $\pi$ pulse,
\begin{align}\label{eq:Evolved Density Matrices}
    \rho_1(\omega) &=  U_{\pi/2} \rho_0 U_{\pi/2}^\dag
    \\
    \rho_2(\omega) &= U_{\mathrm{free}}(\tau_1) \rho_1(\omega) U_{\mathrm{free}}^\dag(\tau_1)
    \\
    \rho_3(\omega) &= U_{\pi} \rho_2(\omega) U_{\pi}^\dag.
\end{align}
The phase accumulated from the $\pi/2$ pulse is 
\begin{equation}\label{eq:Pi2 Accumulated Phase}
    \phi_{\pi/2}(\omega)= \arg \left( \mathrm{tr} (\rho_1(\omega) S_+^\dag) \right)
\end{equation}
and that accumulated from the $\pi$ pulse is
\begin{equation}\label{eq:Pi Accumulated Phase}
    \phi_{\pi}(\omega)= \arg \left( \mathrm{tr} (\rho_3(\omega) S_+^\dagger) \right) - \arg \left( \mathrm{tr} (\rho_2(\omega) S_+^\dagger) \right).
\end{equation}
Unwrapping these phases about the center frequency, the effective evolution times are defined as \cite{kallies2018cooperative}
\begin{align}\label{eq:Effective Time Evolutions}
    \tau_{\mathrm{eff},\pi/2}(\omega) &= \frac{\phi_{\pi/2}(\omega) - \phi_{\pi/2}(0)}{\omega}
    \\
    \tau_{\mathrm{eff},\pi}(\omega) &= \frac{\phi_{\pi}(\omega) - \phi_{\pi}(0)}{\omega}.
\end{align}
(The values of $\tau_{\mathrm{eff},\pi/2}(0)$ and $\tau_{\mathrm{eff},\pi}(0)$ are obtained via interpolation.) This is how long each spin packet would have to freely evolve following an ideal, instantaneous rotation in order to obtain the same final dispersion as produced by the actual rotation. For an echo to form, one must have a linear total phase dispersion (with negative slope) after the $\pi$ rotation. This means any nonlinear phase dispersion acquired from the $\pi/2$ rotation must be canceled by the nonlinear phase dispersion acquired from the $\pi$ rotation. 

In other words, since all the spin packet phases are flipped by the $\pi$ rotation, echo formation requires $\tau_{\mathrm{eff},\pi/2}(\omega) - \tau_{\mathrm{eff},\pi}(\omega)$ to be a constant \cite{kallies2018cooperative}.
Figure \ref{fig:Fig7} shows that this is indeed the case for the FS, DT, and NN pulses, which have nonlinear phase dispersions for the two pulses but cooperate such that their nonlinear parts mutually cancel. Note that the difference $\tau_{\mathrm{eff},\pi/2}(\omega) - \tau_{\mathrm{eff},\pi}(\omega) = t_{\mathrm{refocus}} - \tau_1$. For example, in this case, the FS pulse has $\tau_{\mathrm{eff},\pi/2}(\omega) - \tau_{\mathrm{eff},\pi}(\omega) =$ $-$10 ns and so refocuses at $90$ ns while the NN pulse has $\tau_{\mathrm{eff},\pi/2}(\omega) - \tau_{\mathrm{eff},\pi}(\omega) = 39$ ns and refocuses at $139$ ns. The difference $\tau_{\mathrm{eff},\pi/2}(\omega) - \tau_{\mathrm{eff},\pi}(\omega)$ is shown in Fig.~S4 of the Supplementary Material \cite{sup}. 

\begin{figure}
\includegraphics{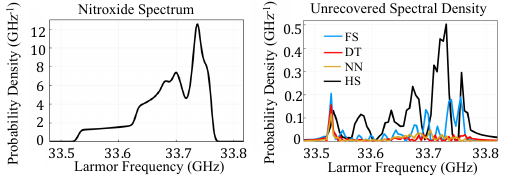}
\caption{Left: an exemplary nitroxide EPR spectrum. Right: the difference between the exemplary nitroxide spectrum and the Fourier transform of the resulting echoes formed by the Fourier Series (FS), discrete time (DT), neural network (NN), and hyperbolic secant (HS) optimized pulse sequences.}
\label{fig:Fig8}
\end{figure}

As a specific example, in Fig.~\ref{fig:Fig8} we use the spectral distribution of a solid-state dilute disordered sample of a nitroxide radical with a bandwidth of about $240$ MHz, shown in the left panel. Because the bandwidth of this distribution is slightly more than the optimized bandwidth of $\delta/2\pi = 240$ MHz, the chosen value of the pulse carrier frequency for each parameterization was also optimized in order to achieve maximum echo height. In the right panel we plot the differences between the actual spectrum and the spectra one would recover from the Fourier transforms of the echoes (note the different scale compared to the left panel). The FS, DT, and NN parameterizations result in less error, particularly around the spectral maximum. All parameterizations have some unrecovered spectral density towards the lower edge of the spectrum due to the bandwidth of the exemplary nitroxide spectrum being slightly larger than the optimized bandwidth.

These optimized pulse shapes have not yet been experimentally implemented. To record the echo generated by these pulses with sufficient fidelity, a receiver of sufficient bandwidth ($>$ 240 MHz) is required. Higher-bandwidth detection systems with up to 1 GHz of bandwidth have become available commercially recently (Bruker SpecJet 3 and VideoAmp 3), so the presented approach is timely and feasible. Also, non-commercial wideband receiver systems have been built \cite{doll_wideband_2017}.

\section{Conclusion}\label{sec:conclusion}

In this paper, we have shown that it is possible to obtain about a $10\%$ improvement of the Hahn echo amplitude over an optimal, generalized hyperbolic secant KBB sequence by optimizing nonlinear amplitude-limited Fourier series, discrete time series, or neural network parameterized pulses. With these parameterized pulses, we find a 1:1 pulse length ratio is favorable in the presence of power constraints because it allows more energy to be allotted to the $\pi$ pulse while the pulse shaping is still able to maintain the refocusing. Interestingly, the optimization landscape for this type of problem has many equivalent maxima, all of which involve the $\pi/2$ and $\pi$ pulses cooperatively working together to compensate phase dispersions in each other. We find no marked differences among the three parameterizations. Furthermore, we have demonstrated that nonlinear effects due to amplifier compression and resonator transfer can be included in the optimization workflow, allowing for the usage of the full power of an amplifier, including its nonlinear region. These results demonstrate a pathway towards optimal broadband spectral acquisition with constrained power.

\begin{acknowledgements}
This research was in part supported by The Under Secretary of Defense for Research and Engineering (OUSD/R\&E) National Defense Education Program (NDEP)/BA-1, Basic Research (E.~R.~L.), NSF CHE-2154302 (S.~S.), NSF PHY-1915064 (J.~P.~K.), and is based upon work supported by the National Science Foundation under the STC IMOD Grant No.~2019444 (S.~S. and J.~P.~K.). 

This work used the Advanced Research Computing at Hopkins (ARCH) core facility (rockfish.jhu.edu) through allocation XSEDE-PHY230011 from the Advanced Cyberinfrastructure Coordination Ecosystem: Services \& Support (ACCESS) program \cite{ACCESS}, which is supported by National Science Foundation grants \#2138259, \#2138286, \#2138307, \#2137603, and \#2138296. This work also used the UMBC High Performance Computing Facility (HPCF), which is supported by the NSF through the MRI program (Grant Nos.~CNS-0821258, CNS-1228778, OAC-1726023, and CNS-1920079) and the SCREMS program (Grant No.~DMS-0821311), with additional substantial support from the University of Maryland, Baltimore County (UMBC).
\end{acknowledgements}

\bibliography{references}

\end{document}


\title{Supplementary Material: Optimizing EPR pulses for broadband excitation and refocusing}
\author{Eric R. Lowe}
\email{elowe2@umbc.edu}
\affiliation{Department of Physics, University of Maryland Baltimore County, Baltimore, MD 21250, USA}
\author{Stefan Stoll}
\email{stst@uw.edu}
\affiliation{Department of Chemistry, University of Washington, Seattle, WA 98195, USA}
\author{J.~P.~Kestner}
\email{jkestner@umbc.edu}
\affiliation{Department of Physics, University of Maryland Baltimore County, Baltimore, MD 21250, USA}
\email[]{stst@uw.edu, jkestner@umbc.edu}

{
\let\clearpage\relax
\maketitle
}
\vspace{-1cm}

\section{$I/Q$ Optimized Shapes}\label{sec:I/Q}
In the results section of the main manuscript, the optimized pulse shapes which the spins `see' are plotted for the three parameterizations: Fourier series (FS), discrete time (DT), and neural network (NN). These are the pulse shapes after passing through the entire distortion chain. Below in Fig.~\ref{fig:PlotPulse} we plot the $I/Q$ shapes before passing through the distortion chain for these optimized parameterizations.

From Fig.~\ref{fig:PlotPulse}, it is clear that for all $\pi$ pulses, the optimized shapes are using maximum amplitude for almost the entire duration of the pulse, toggling between an amplitude of $1$ and $-1$. This is also a feature of any optimized pulse solution regardless of parameterization as the $\pi$ pulse requires more power than the $\pi/2$ pulse. Looking at the $\pi/2$ pulses, the NN does not need to use maximum amplitude for the duration of the pulse while the FS and DT use more power. This is not a feature of the NN parameterization, it is a consequence that the $\pi/2$ pulse has more flexibility in equivalent maxima.

\begin{figure*}[h!]
\includegraphics{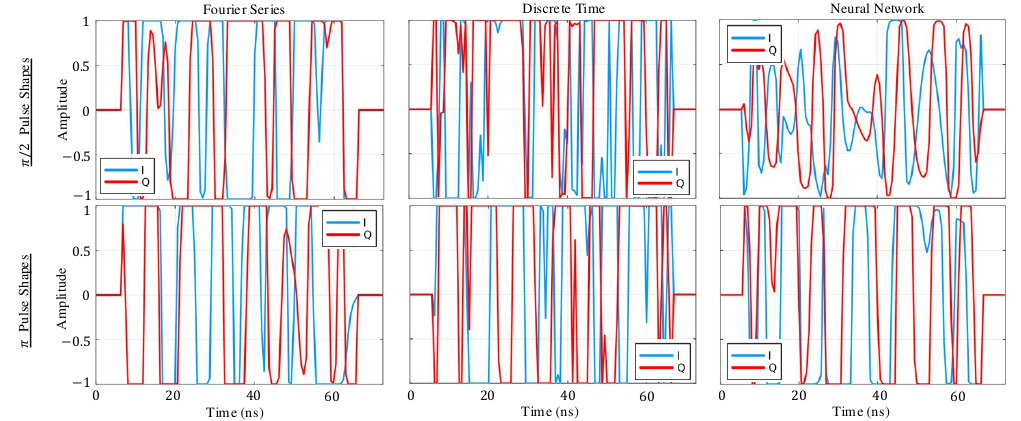}
\caption{Plots of the $I/Q$ shapes before passing through the distortion chain. The top row of pulse shapes are for the $\pi/2$ pulses while the bottom row of pulse shapes area for the $\pi$ pulses. The three columns in order from left to right represent the FS, DT, and NN pulse shapes further analyzed in the results section of the main manuscript.}
\label{fig:PlotPulse}
\end{figure*}

\section{$M_x$,$M_y$,$M_z$}\label{sec:MxMyMz}
\begin{figure*}[!h]
\includegraphics{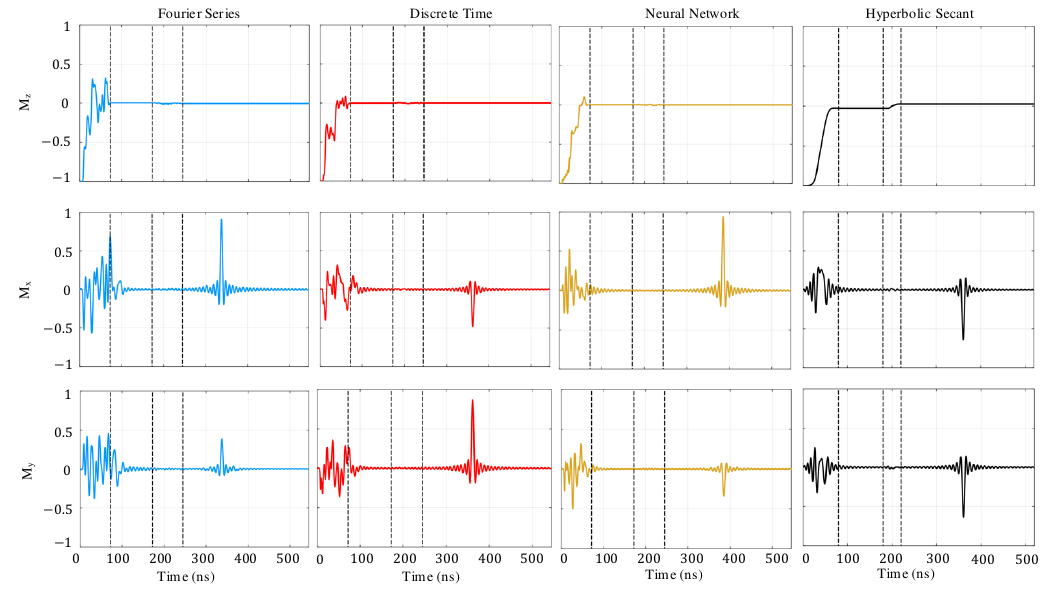}
\caption{Plots of the magnetization, $M_x$,$M_y$,$M_z$, for each of the $4$ optimized parameterizations. The evolution leading up to the first dashed line represents the evolution of the $\pi/2$ pulse. The next leading up to the second dashed line is the free evolution period between pulses. The next evolution to the final dahsed line is the evolution of the $\pi$ pulse. The final space is the free evolution period after both pulses.}
\label{fig:MxMyMz}
\end{figure*}

The total magnetization in $x$,$y$, and $z$ are all plotted in Fig.~\ref{fig:MxMyMz} during the pulse sequences for all $4$ optimized sequences. For the hyperbolic secant (HS) pulse parameterization, the $\pi/2$ pulse evolution in the $M_z$ plot shows a straightforward excitation of all spins in a single sweep. The numerically designed pulses are less transparent in the spin dynamics, but are all able to obtain a $\pi/2$ excitation regardless. All projections, $M_x$,$M_y$,$M_z$, remain at $0$ during the $\pi$ pulse as spins are dispersed evenly along the $xy$-plane and so their magnetizations average out to $0$.

\begin{figure}[h!]
\includegraphics{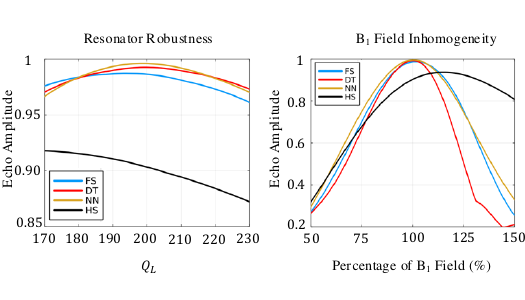}
\caption{On the left, a plot of echo amplitude versus different $Q_\mathrm{L}$ values of the resonator. All pulses were optimized for a value of $Q_L=200$. On the right, a plot of echo amplitude versus percentage of $B_1$ field. Again, all pulses were optimized for a single value of $100\%$.}
\label{fig:ResonatorRobustness}
\end{figure}
\section{Resonator and $B_1$ field Robustness}\label{sec:ResonatorRobustness}
In Fig.~\ref{fig:ResonatorRobustness}, we plot the robustness of each of the $4$ optimized pulse parameterizations to $Q_L$ of the resonator, from equation (11) of the main manuscript. While each of the parameterizations were optimized for a value $Q_L=200$, we see that a lower value of $Q_L$ actually improves the echo amplitude for the HS pulse sequence. This is because the resonator profile for a lower value of $Q_L$ inherently has a wider $3$ dB bandwidth, meaning that more power can reach the spins for a given sweep across the bandwidth. As mentioned in the results section of the main manuscript, the issue with the HS pulse shape is the $\pi$ pulse's inability to fully flip every spin across the bandwidth, and so a lower value $Q_L$ counteracts this. The $3$ numerically designed pulse shapes show an equal robustness to a change in $Q_L$ and are relatively optimal for $Q_L = 200$.

In the plot of $B_1$ Field Inhomogeneity, we plot the echo amplitude versus $B_1$ field inhomogeneity. Similar to that of the resonator robustness plot, we find the HS is maximal for a larger $B_1$ field. This is again because the $\pi$ pulse requires more power and a larger $B_1$ field accomplishes this. The FS and NN pulse shapes are equally robust to $B_1$ field inhomogeneity and the DT pulse shape is less robust at higher $B_1$ fields.

\section{Effective Time Evolution}
\begin{figure}[h]
\includegraphics{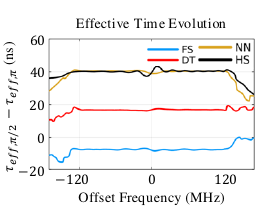}
\caption{Difference in the effective time evolution of the $\pi/2$ and $\pi$ pulse calculated from Eqs.(42) and (48) plotted as a function of offset frequency for each of the $4$ optimized parameterizations.}
\label{fig:TimeEvolution}
\end{figure}

In the main manuscript in Fig. 7, we plot the effective time evolution for the $\pi/2$ and $\pi$ pulses for each of the $4$ optimized pulse shapes. Here, in Fig.~\ref{fig:TimeEvolution}, we plot the difference of these two curves. This represents the time of free evolution until the peak of the formed echo plus the free evolution period in between pulses, that being $100$ ns. From this plot we see that both the HS and NN effective evolution times are around $40$ in between the designated bandwidth of $120$ MHz, which corresponds to an echo formation that is $140$ ns after the end of their respective $\pi$ pulses. The DT and FS pulse sequences refocus their spins earlier and so their respective time evolution is less. From the plot, it is also clear that there is some dephasing for the HS pulse sequence near the carrier frequency. 

